\documentclass[sigconf,screen, authorversion, nonacm]{acmart}

\setcopyright{acmlicensed}
\acmPrice{15.00}
\acmDOI{10.1145/3611643.3616255}
\acmYear{2023}
\copyrightyear{2023}
\acmSubmissionID{fse23main-p126-p}
\acmISBN{979-8-4007-0327-0/23/12}
\acmConference[ESEC/FSE '23]{Proceedings of the 31st ACM Joint European Software Engineering Conference and Symposium on the Foundations of Software Engineering}{December 3--9, 2023}{San Francisco, CA, USA}
\acmBooktitle{Proceedings of the 31st ACM Joint European Software Engineering Conference and Symposium on the Foundations of Software Engineering (ESEC/FSE '23), December 3--9, 2023, San Francisco, CA, USA}
\received{2023-02-02}
\received[accepted]{2023-07-27}

\ccsdesc[500]{Software and its engineering~Software testing and debugging}

\usepackage{balance}
\usepackage{pifont}
\usepackage[ruled,vlined]{algorithm2e}
\usepackage{microtype}
\usepackage{libertine}

\RequirePackage{xcolor}
\RequirePackage{hyperref}
\definecolor{orcidlogocol}{HTML}{A6CE39}

\DeclareRobustCommand\orcidlink[1]{%
  \href{https://orcid.org/#1}{%
    \texorpdfstring{\includegraphics[height=1em]{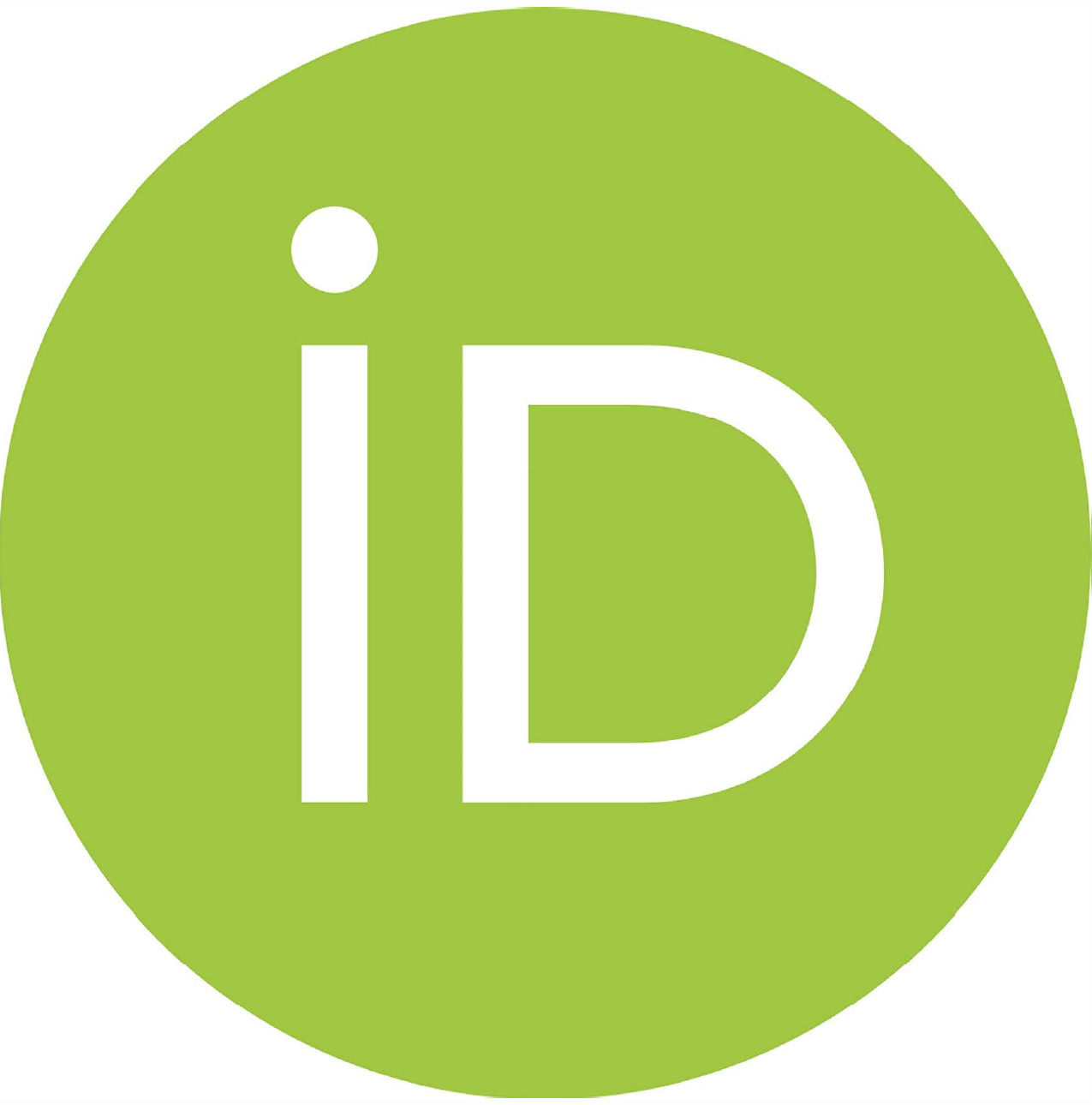}}{ORCID}%
  }%
}

\newcommand{\Ericsson}{Ericsson\xspace}
\newcommand{\TestAll}{\textit{TestAll}\xspace}
\newcommand{\BatchStopFour}{\textit{BatchStop4}\xspace}

\newcommand{\ConstantBatching}{\textit{ConstantBatching}\xspace}
\newcommand{\BatchTwo}{\textit{Batch2}\xspace}
\newcommand{\BatchFour}{\textit{Batch4}\xspace}
\newcommand{\DynamicBatching}{\textit{BatchAll}\xspace}
\newcommand{\TestCaseBatching}{\textit{TestCaseBatching}\xspace}
\newcommand{\ExecutionReduction}{\textsc{ExecutionReduction}\xspace}

\newcommand{\mr}{{machine}\xspace}
\newcommand{\mrs}{{machines}\xspace}

\newcommand{\etal}{\hbox{\emph{et al.}}\xspace}
\newcommand{\eg}{\hbox{\emph{e.g.,}}\xspace}
\newcommand{\ie}{\hbox{\emph{i.e.}}\xspace}

\begin{document}

\title{Accelerating Continuous Integration with Parallel Batch Testing}

\author{Emad Fallahzadeh}
\orcid{0009-0005-5024-4868}
\affiliation{%
  \institution{Concordia University}
  \department{Department of Computer Science \\and Software Engineering}
  \city{Montr{\'e}al}
  \state{Qu{\'e}bec}
  \country{Canada}
  }
\email{e_falla@encs.concordia.ca}

\author{Amir Hossein Bavand}
\orcid{0000-0003-4913-6729}
\affiliation{%
  \institution{Concordia University}
  \department{Department of Computer Science \\and Software Engineering}
  \city{Montr{\'e}al}
  \state{Qu{\'e}bec}
  \country{Canada}
  }
\email{a\_bavand@encs.concordia.ca}

\author{Peter C. Rigby}
\orcid{0000-0003-1137-4297}
\affiliation{%
  \institution{Concordia University}
  \department{Department of Computer Science \\and Software Engineering}
  \city{Montr{\'e}al}
  \state{Qu{\'e}bec}
  \country{Canada}
  }
\email{peter.rigby@concordia.ca}

\thanks{© 2023 Copyright held by the owners/authors. Publishing rights licensed to ACM. This is the author's accepted version of the work. It is posted here for your personal use. Not for redistribution. The definitive version will be published in Proceedings of the 31st ACM Joint European Software Engineering Conference and Symposium on the Foundations of Software Engineering (ESEC/FSE '23), December 3--9, 2023, San Francisco, CA, USA, \url{https://doi.org/10.1145/3611643.3616255}.}

\keywords{Batch Testing, Parallel, Large-Scale, Feedback, Execution Reduction}

\begin{abstract}
Continuous integration at scale is costly but essential to software development. 
Various test optimization techniques including test selection and prioritization aim to reduce the cost. 
Test batching is an effective alternative, but overlooked technique.
This study evaluates parallelization's effect by adjusting machine count for test batching and introduces two novel approaches.

We establish \TestAll as a baseline to study the impact of parallelism and machine count on feedback time. We re-evaluate \ConstantBatching and introduce \DynamicBatching, which adapts batch size based on the remaining changes in the queue. We also propose \TestCaseBatching, enabling new builds to join a batch before full test execution, thus speeding up continuous integration. Our evaluations utilize \Ericsson's results and 276 million test outcomes from open-source Chrome, assessing feedback time, execution reduction, and providing access to Chrome project scripts and data.


The results reveal a non-linear impact of test parallelization on feedback time, as each test delay compounds across the entire test queue. \ConstantBatching, with a batch size of 4, utilizes up to 72\% fewer machines to maintain the actual average feedback time and provides a constant execution reduction of up to 75\%. Similarly, \DynamicBatching maintains the actual average feedback time with up to 91\% fewer machines and exhibits variable execution reduction of up to 99\%. \TestCaseBatching holds the line of the actual average feedback time with up to 81\% fewer machines and demonstrates variable execution reduction of up to 67\%.
%
We recommend practitioners use \DynamicBatching and \TestCaseBatching to reduce the required testing machines efficiently. Analyzing historical data to find the threshold where adding more machines has minimal impact on feedback time is also crucial for resource-effective testing.

\end{abstract}   

\maketitle

\section{Introduction}

\label{sec:introduction}


Testing is both time-consuming and resource intensive. 
To reduce both resource consumption and provide earlier feedback, test selection has been widely adopted in the industry and extensively studied~\cite{536955, ENGSTROM201014, chen_using_2011}. Test selection's inherent trade-off is that not all tests are run, and some test failures may be missed.
On the other hand, test prioritization guarantees that all tests will be run, but those that are more likely to reveal faults will be run first, reducing feedback time on test failures, but not reducing the resource usage in testing~\cite{bagherzadeh_reinforcement_2022, zhu_test_2018, 962562}.
In contrast, batch testing which groups builds for testing and bisects on failure is conceptually better than both test selection and prioritization because all the tests are run with less resource consumption and much faster overall feedback times~\cite{najafi_bisecting_2019, beheshtian_software_2021}.
Most changes require similar test sets, and the saving in batch testing is achieved because batching groups changes and tests to reduce the number of redundant test runs among builds. For example, if we are testing four builds in a batch that request the same tests, and the batch passes, we will save three build test executions. However, when a batch fails, a bisection algorithm must be run to identify the culprit build that is causing the failure. Bisection can slow individual builds, but when fewer than 40\% of builds fail, batching and bisection effectively reduce overall feedback time~\cite{beheshtian_software_2021}.


 
Despite its effectiveness, only few studies examine batch testing.
Najafi \etal ~\cite{najafi_bisecting_2019} studied batching and bisection techniques at Ericsson and found that constant batch sizes can reduce the resource usage, \ie execution time, necessary to run all the required tests by up to 42\%.
 Beheshtian \etal ~\cite{beheshtian_software_2021} evaluated open-source Travis torrent projects and proposed \BatchStopFour, which can reduce build test executions on average by 50\%.
 Bavand \etal ~\cite{bavand_mining_2021} used a dynamic batch size approach using the historical failure rate of the projects and reduced the execution time by about 47\% against their baseline.
Previous batch testing works make three assumptions that are unrealistic on large software systems. First, they do not run tests in parallel and implicitly use a single machine. Second, after the failure of a batch, they rerun all tests. This is inefficient because we know which tests failed on a batch, and we do not need to bisect and re-run the passing tests. Third, previous researchers only focus on reducing resource consumption and do not investigate the feedback time outcomes for different batching algorithms.



In this study, we address the limitations of prior work, and study systems with a much larger scale of CI builds and tests: \Ericsson and Chrome.
To understand the impact of parallelization on testing, we replicate the \TestAll algorithm which simply runs all tests without applying any batching technique, but unlike prior work, we vary the number of machines available for testing and run tests in parallel. Our outcome measures are the feedback time, \ie wall time per test, and total test execution time reduction for all the tests.
We replicate \ConstantBatching from the Beheshtian \etal's study \cite{beheshtian_software_2021} which uses a constant batch size for batching, and we introduce two novel \DynamicBatching and \TestCaseBatching techniques.
\DynamicBatching algorithm adapts the batch size to the number of remaining changes inside the queue to batch them all at each time.
\TestCaseBatching approach works the same as \DynamicBatching except that it also accepts new changes while running the batch.
We adopt different numbers of machines on these batching algorithms and evaluate their performance by using feedback time and execution reduction.

We provide results for the following research questions:

\textit{RQ1: How does parallelization affect the feedback time performance of \TestAll with varying numbers of machines?}

\textit{RQ2: How effective is \ConstantBatching in terms of feedback time and execution reduction when executed in parallel with varying numbers of machines?}

\textit{RQ3: How effective is \DynamicBatching in terms of feedback time and execution reduction when executed in parallel with varying numbers of machines?}

\textit{RQ4: How effective is \TestCaseBatching in terms of feedback time and execution reduction when executed in parallel with varying numbers of machines?}        

Our major contributions to this study are as follows.
\begin{itemize}
    \item \textbf{Parallelism in batching:} we study the impact of parallelization on testing in general and batching.
    \item \textbf{Datasets:} we use two large-scale \Ericsson and Chrome datasets which run millions of tests per day and are suitable to run our batching algorithms and apply parallelism.
    \item \textbf{Simulation:} we replicate \TestAll and \ConstantBatching algorithms and introduce two novel \DynamicBatching and \TestCaseBatching approaches.
    \item \textbf{Outcome measure:} we use feedback time and execution reduction measures to evaluate the performance of batching algorithms.
    \item \textbf{Results:} we reach the following conclusions.
    \begin{enumerate}
        \item The impact of parallelism on the performance testing algorithms exhibits a non-linear relationship.
        \item \ConstantBatching with a batch size of 4, utilizes up to 72\% fewer machines to maintain the actual average feedback time and provides up to 75\% in execution reduction, regardless of the number of machines utilized. 
        \item \DynamicBatching maintains the actual average feedback time with up to 91\% fewer machines and exhibits variable execution reduction of up to 99\% depending on machine utilization and the dataset.
        \item \TestCaseBatching holds the line of the actual average feedback time with up to 81\% fewer machines and demonstrates variable execution reduction of up to 67\% depending on machine utilization and the dataset.
    \end{enumerate}
\end{itemize}



\section{Background and Methodology}
\label{sec_background_method}

We study a proprietary project at \Ericsson and the Google-led open source project Chrome.
The project we examine at \Ericsson tests the software that runs on cellular base stations.
In this context, the \mrs used for testing are extremely expensive and limited in number.
\Ericsson spends millions on testing infrastructure and still needs to batch tests in order to test all the changes. Test resources are scarce, and managers discuss the tradeoff of buying new test machine resources for having slower feedback or removing more tests through selection and risking failures slipping through. 


Testing at \Ericsson involves a multistage process, including various testing levels from unit tests to integration tests, before changes are integrated into the released product. Our study focuses on confidence levels 2 and 3, primarily consisting of integration tests, which significantly contribute to the overall testing costs at \Ericsson compared to other levels like unit testing. Thus, we capture the test results, execution time, and change timestamps for each test case during integration testing to gain insights and optimize this critical stage.


To generate our sample dataset, we evaluate the integration testing of a project at \Ericsson. 
We capture the test results for the period of six weeks from January to February 2021. 
For this duration, we observe over 11,000 changes.
For confidentiality reasons, we do not disclose other data or aspects of the testing process and use the data to simulate batching scenarios where machines are very expensive and highly utilized leading to strong resource constraints.

\textbf{Chrome} is one of the most popular browsers in the world, and it is representative of a large-scale project. 
There are millions of tests run each day, and there are a massive number of builds and tests running in parallel.
The testing process used by Chrome was described in detail by Fallahzadeh \etal \cite{fallahzadeh_impact_2022}, we summarize briefly below.
A change list, \ie pull request, is committed to the Gerrit code review tool for revision. 
Reviewers may suggest changes to the change list to improve the code or when they find issues. If the change is satisfactory to the reviewers, it will be sent for testing on the try bot builders. 
After being approved by the builders, the change is merged into the main Chrome repository.

For the Google Chrome case, we use the Chrome test results published by Fallahzadeh \etal \cite{fallahzadeh_impact_2022}.
This publicly available dataset is captured by calling the Gerrit code review APIs from Chrome. 
This dataset consists of 276 million test cases for the month of January 2021 for Chrome.
These test runs are for 9,524 changes across 49,932 distinct test suites for an average of 8.9 million test case runs per day. 
The rate of failures for the builds in this project is about 8.5\%.
We describe how this data is used in our simulation below. 

For both \Ericsson and Chrome, the data of interest are the test id, test name, build id, build start time, build end time, status, final result, and test duration.
We use these attributes to implement our various batching algorithms which will be discussed in the following. This methodology can be applied to any project that collects this basic data. We release our scripts and data for Chrome in the supplemental material in an anonymized replication package.


\subsection{Simulation Method and Outcome Measures}


In this study, we use the record of real historical test runs and only vary the number of \mrs to evaluate different batching algorithms. We do {\it not} re-run any tests. Instead, the process involves determining tests for a batch based on the historical test results of the included builds, selecting the maximum execution time among the corresponding builds for each test to capture the worst-case execution time. The simulated batch is then dispatched across available machines, optimizing test allocation for parallelism and resource utilization, considering the varying number of machines in different scenarios. During the simulation, we assume that a test will fail in the simulated batch if it failed in any of the builds included in the batch, and a test is considered to pass in the simulated batch only if it passed in all the builds. This approach ensures an accurate reflection of test behavior and outcomes in the simulation. We preserve the order of test runs and discuss the limitations of simulation in Section \ref{sec_threats}. 

The data necessary to conduct the simulation is not company specific and the simulations can be applied to the test results of the other projects as well.
The data includes the code change under test, the time the change was available for testing, the requested tests for the change, and the duration that each test took to run. 
The changes are then queued based on their arrival time and simulated using the batching algorithm, \ie how efficiently would we have been able to process the same changes and requested tests?

\textbf{Feedback time:} One of the most important factors in designing a continuous integration testing infrastructure is giving fast feedback on test outcomes for each change. The time between committing a change and receiving all test verdicts is defined as feedback time. 

\begin{equation}
\label{eq_CorrectedReduction}
{\text{FeedbackTime}} = \text{Time}_{\text{TestVerdicts}}-\text{Time}_{\text{Commit}}
\end{equation}

For example, if a developer commits a change at 9 am and receives the feedback that the tests passed successfully on that change at 10 am, the feedback time will be 1 hour for that change. In contrast, if the change was queued for 1 hour, then the feedback time would be 2 hours, a doubling in feedback time. In the examples, we use a unit $T$ to represent time, but in the study, we use the actual time each test takes to run. The time a test will be queued depends on the available resources and batching algorithm. 

To contrast batching algorithms, we use the \textit{AvgFeedback} as the sum of the feedback times for each change in the project divided by the total number of changes for the project. The equation below shows the \textit{AvgFeedback} for batching algorithm $A$ across $C$ changes with $m$ \mrs.

\begin{equation}
\text{AvgFeedback}_m(A)=\frac{\sum_{c=1}^{C}{\text{Feedbacktime}_c(A, m)}}{C}
\label{avg_feedback_time_equation}
\end{equation}

To contrast batching algorithms, we calculate the \textit{FeedbackReduction} as the percentage decrease in \textit{AvgFeedback} of  batching algorithm, A1, compared to the \textit{AvgFeedback} of the algorithm, A2, with $m$ \mrs.

\begin{equation}
\text{FeedbackReduction}_m(A1, A2)=(1-\frac{\text{AvgFeedback}_{m}(A1)}{\text{AvgFeedback}_{m}(A2)}) * 100
\label{feedback_reduction_equation}
\end{equation}

\textbf{Execution Reduction:}
The \ExecutionReduction metric quantifies the performance of different batching algorithms in terms of saving execution time with varying numbers of machines. It is calculated using Equation \ref{eq:time_saved}.

\begin{multline}
    \text{\ExecutionReduction}_m(A) = \\ 1 - \frac{\sum_{t=1}^{K} \text{Test Execution Time (A, m)}}{\sum_{t=1}^{N} \text{Test Execution Time (\TestAll)}}
    \label{eq:time_saved}
\end{multline}

This formula calculates the execution reduction in test executions achieved by a batching algorithm (A) with m machines compared to running all tests in their original order (\TestAll). Here, $K$ represents the number of tests executed by the batching algorithm (A) with m machines, and $N$ is the total number of tests executed by \TestAll. The formula calculates the percentage of time required for running the tests using approach A with m machines relative to \TestAll. Subtracting this percentage from 1 provides the percentage of execution time that was reduced. For example, if approach A with m machines takes 40\% less time to run tests than \TestAll, the \ExecutionReduction by A with m machines is calculated as 60\%, indicating that approach A with m machines saves 60\% of the time in test execution compared to \TestAll.

By using this formula as a metric, we can compare the performance of different batching algorithms and determine their efficiency in saving time during test execution.

\textbf{\textit{Simulation Setup and Plateau Thresholds.}}
To evaluate the batching algorithms' performance in terms of feedback time, we compare them to the actual average feedback time as the baseline. Due to the complexity and unavailability of the exact number of machines required to achieve the actual average feedback time in both \Ericsson and Chrome, we use the baseline number of machines needed to maintain the actual average feedback time for \TestAll. Our simulations vary the available resources (\mrs) from 1 to 9 for \Ericsson and from 1 to 375 for Chrome, where the average feedback time for all algorithms reaches a plateau.

Determining a plateau often involves the expertise of domain experts who suggest appropriate thresholds. For \Ericsson, we establish a threshold of 6 percent improvement from the actual feedback baseline, indicating that beyond this point, the improvement in feedback time becomes insignificant with a unit increase of 1 machine. In the case of Chrome, the threshold is set at 2 percent improvement from the baseline, considering a unit increase of 25 machines. These thresholds are carefully determined to account for the specific requirements and trade-offs in each domain. \Ericsson, with its high-cost machines, necessitates a more stringent plateau criterion, while Chrome, benefiting from machine farms, adopts a lower threshold to optimize performance.

\section{Batching Algorithms}
\label{sec_batching}

Although parallel testing can help to reduce the feedback time, even at large companies using farms of servers to run tests in parallel, they still need batch changes to further reduce resources~\cite{ziftci_who_2017}. In the following, we describe the definitions and formulations for each batching algorithm.

\subsection{\TestAll}

Ideally, each change would be tested immediately and in isolation, running all the requested tests independently of other changes. This approach works well on small projects that are not resource constrained. 
The feedback time for each change varies and depends on the time that a change waits in the queue.

\begin{figure}
	\centering
	\includegraphics[width=\columnwidth]{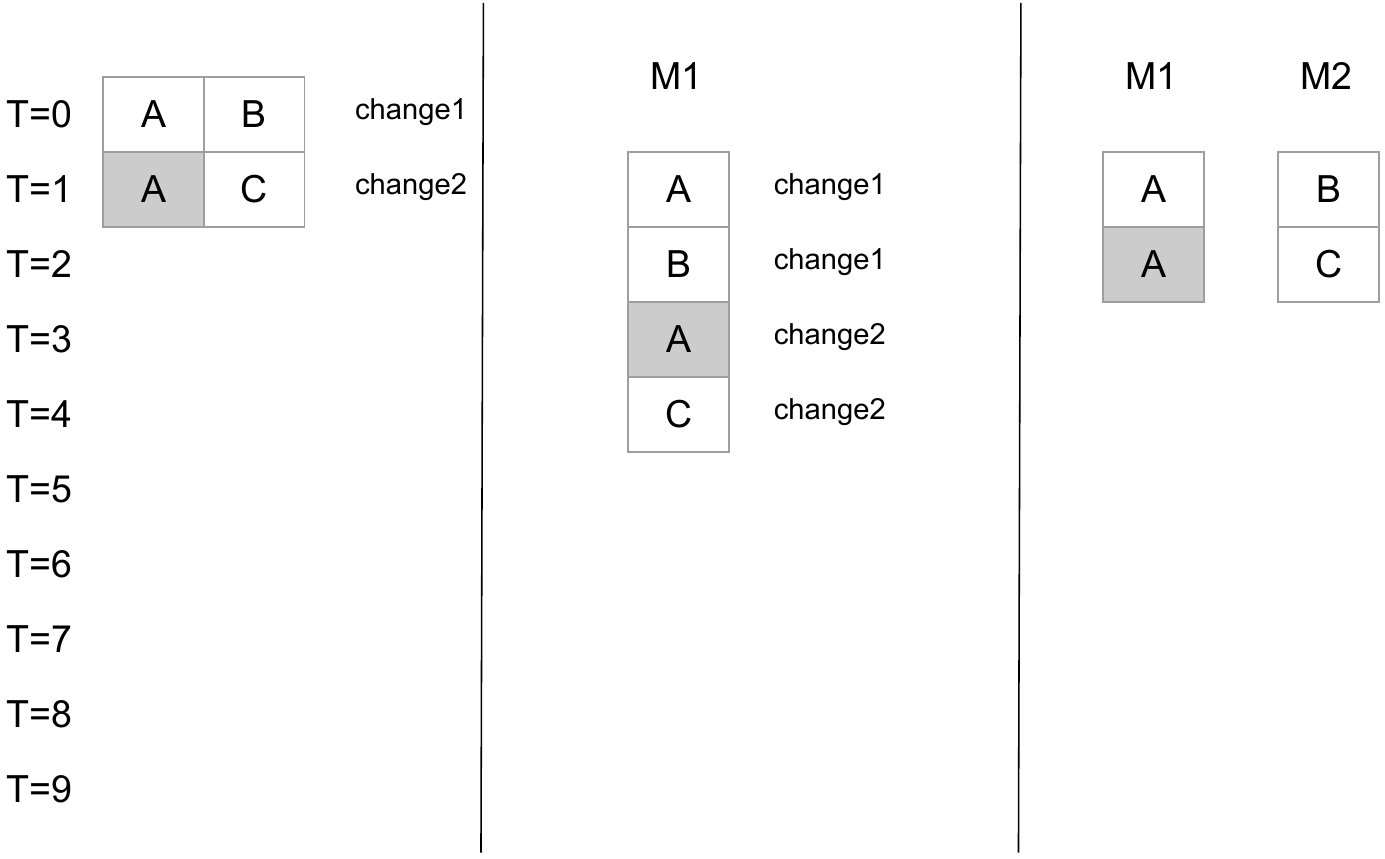}
	\caption[An example of \TestAll methodology]{
 A sample of \TestAll algorithm.
In this example, there are two subsequent changes and their requested tests A, B, and C.
The sequence of events is displayed by T, and the machines are shown by M.
	}
	\label{fig:testAll}
\end{figure}

Figure \ref{fig:testAll} describes the testing process for \TestAll algorithm with 1 and 2 \mrs scenarios. There are two changes that arrive at $T=0$ and $T=1$ times respectively. When there is only one \mr available, all tests are running on a single machine and subsequent changes have to wait in a queue. At the time $T=2$, the tests belonging to the first change are executed with a feedback time equal to 2 units of time. Test execution for change 2 is finished at $T=4$ resulting in the feedback time of 3 units of time.
In the 2 \mrs scenario, the tests belonging to the changes are distributed between the two \mrs. At the time $T=1$ both tests A and B for change 1 are executed by the machines M1 and M2, leading to 1 unit of feedback time. At the time $T=2$, tests are run for all changes, making the average feedback time of 1 in comparison with the previous average feedback time of 2.5 units of time. 

\subsection{Batch Testing and \BatchStopFour}

\TestAll is expensive and sometimes infeasible at large companies, \eg Ericsson~\cite{najafi_bisecting_2019} or Google~\cite{ziftci_who_2017}. Instead of testing every change individually, we can combine multiple changes and run the union of their requested tests in a batch. If a batch passes, we save resources and provide feedback more quickly. However, if a test fails, we need to find the culprit change(s) that are responsible for the failure. If the intersection of the requested tests for the batched changes is large, and most tests pass, the saving could be substantial. In an extreme example, if we batch 50 changes, and each change requests the same tests when the batch passes we save 49 build test executions. 

\begin{figure}
	\centering
	\includegraphics[width=\columnwidth]{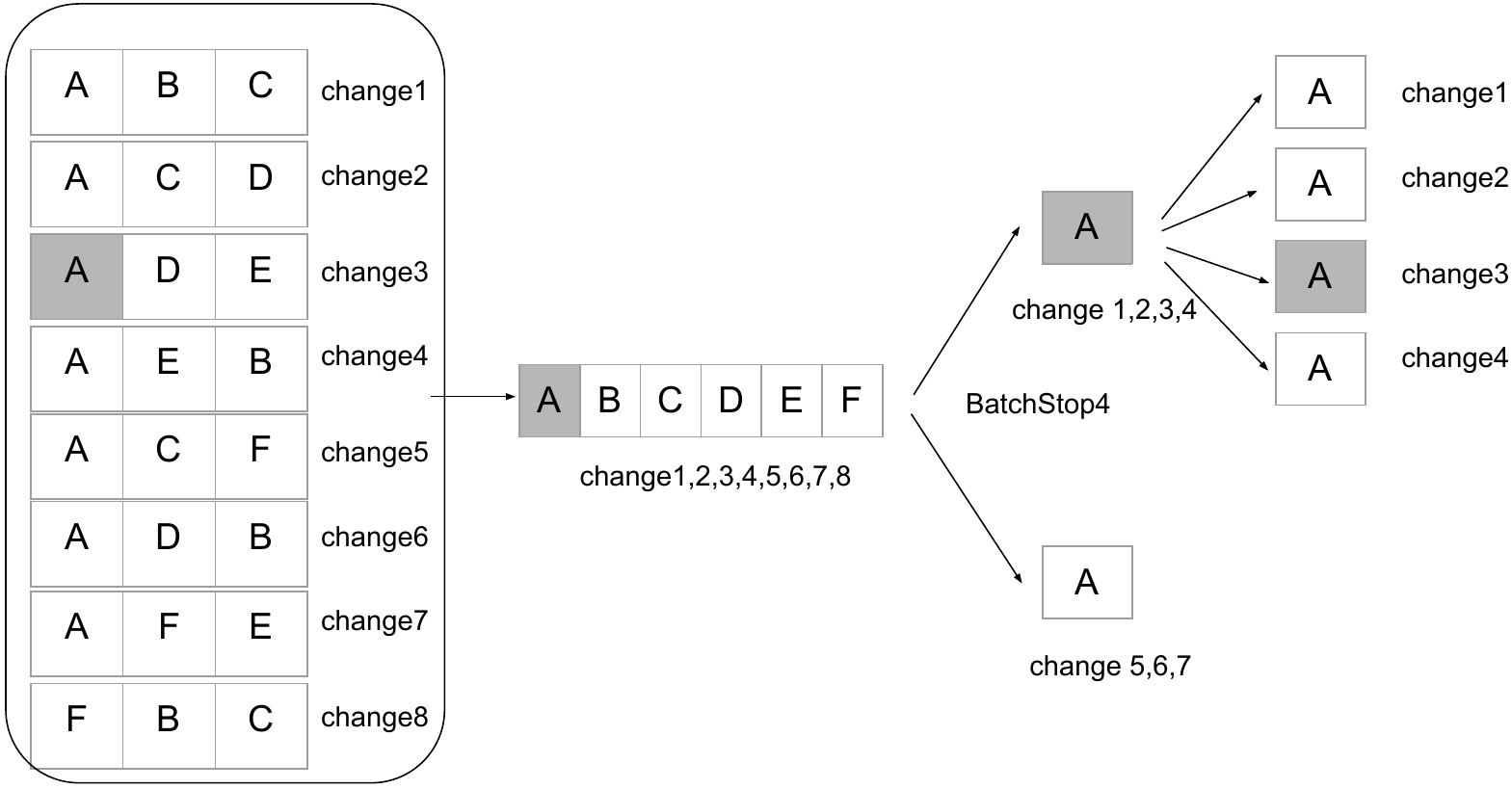}
	\caption[An example of Batching and \BatchStopFour]{
 A sample of batch testing using the \BatchStopFour culprit approach.
In this example, there are eight subsequent changes and their requested tests A, B, C, D, E, and F. 
The failing test A is determined by a gray colour.
}
	\label{fig:Batching}
\end{figure}

However, when the batch fails, we need to find the culprit change(s) responsible for the failure. Out of different culprit-finding approaches like Dorfman and bisection~\cite{dorfman1943, najafi_bisecting_2019}, we use the \BatchStopFour which has been shown both mathematically and empirically to be the top performing approach~\cite{beheshtian_software_2021}. 
Figure \ref{fig:Batching} displays the batching process of 8 changes consisting of different tests with the same execution time. To create the batch, a union of the tests across all the changes is used, which gives 6 distinct tests of A, B, C, D, E, and F to run. Executing the batch fails, leading to the culprit-finding process that requires additional 6 test executions. Since we know that test A failed, we only run this test in the subsequent builds. This ends up running 12 test runs, which is 50\% less than the \TestAll approach that requires 24 test executions. 

\subsection{\ConstantBatching}
\begin{figure}
	\centering
	\includegraphics[width=\columnwidth]{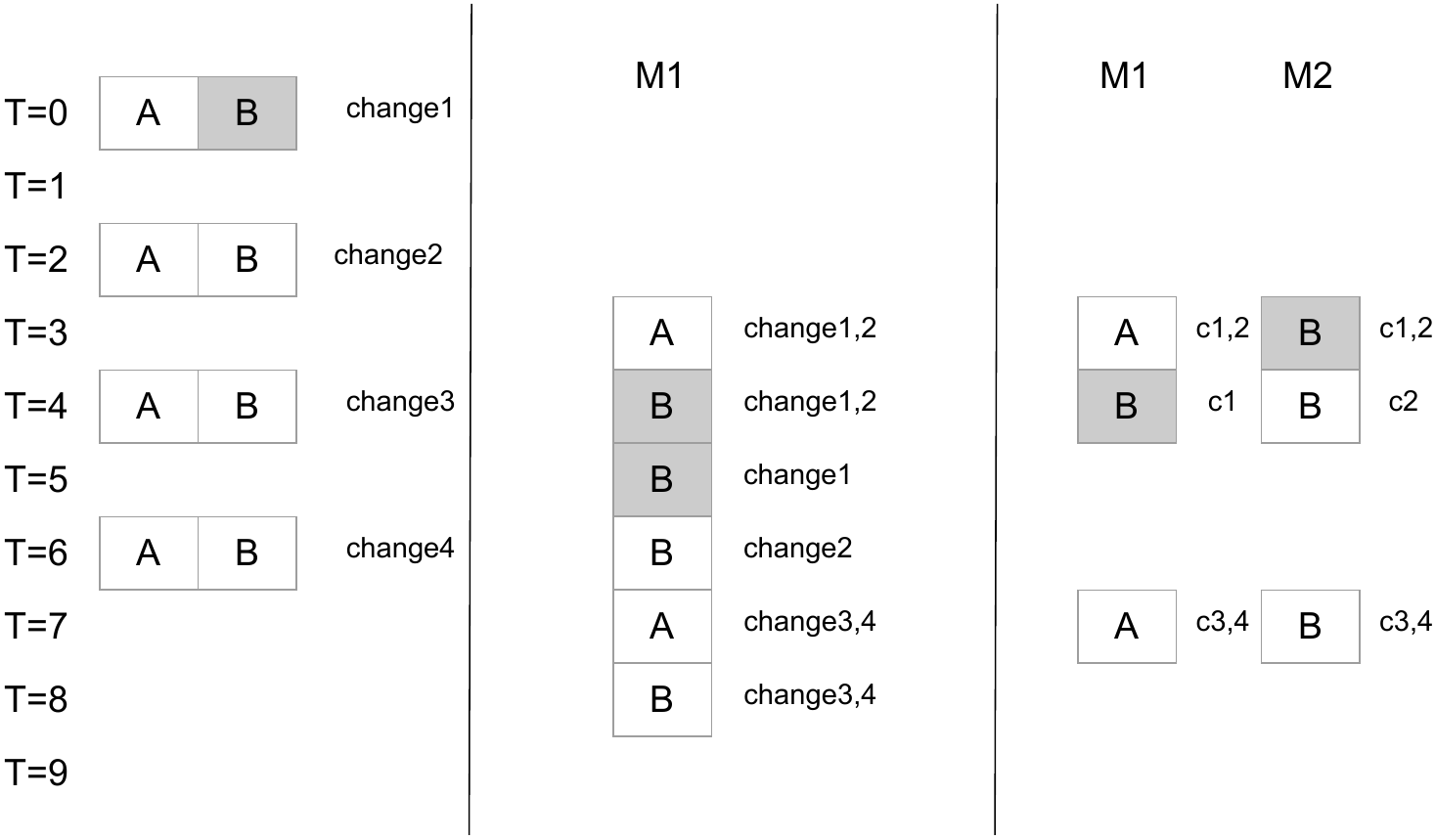}
	\caption[An example of \ConstantBatching]{
 A sample of \ConstantBatching algorithm \ie \BatchTwo.
In this example, there are four subsequent changes and their requested tests A, and B.
The sequence of events is displayed by T, and the machines are shown by M.
 }
	\label{fig:ConstantBatching}
\end{figure}

Prior works have selected a constant batch size for testing~\cite{najafi_bisecting_2019, beheshtian_software_2021}. In the \ConstantBatching technique, we group $n$ changes together and test them in a batch. For example, with $n=8$, we batch every 8 changes together for testing. Figure \ref{fig:ConstantBatching} shows an example of \BatchTwo using a single \mr as well as having two parallel \mrs. There are in total 4 changes. The time of committing the changes is equal to T=0, T=2, T=4, and T=6 respectively. For simplicity, we assume each test execution takes 1 unit of time. 
In a single \mr configuration, the feedback time for each build would be 5, 4, 4, and 2 units of time respectively.
By using two \mrs, the feedback time would be 4, 2, 3, and 1 unit of time respectively.
We can see when using two parallel \mrs, there is a time when \mrs are free and no test has been assigned to them for execution as they have to wait for 2 changes to be available. 
This affects the feedback time for Change 3. 

\subsection{\DynamicBatching}

The assumption of a constant batch size introduces problems. First, the rate of committed  changes varies over time. For example, during the peak of the workday, there may be 1000's more commits than at night. We need to vary the batch size based on the change queue. In \DynamicBatching, when there are resources available, all the waiting changes, are grouped and the union of required tests is run for the batch. When the testing process of the batch finishes and the corresponding resources are free, another batch is created using all the current waiting changes and the resources are allocated to the new batch. 

\begin{figure}
	\centering
	\includegraphics[width=\columnwidth]{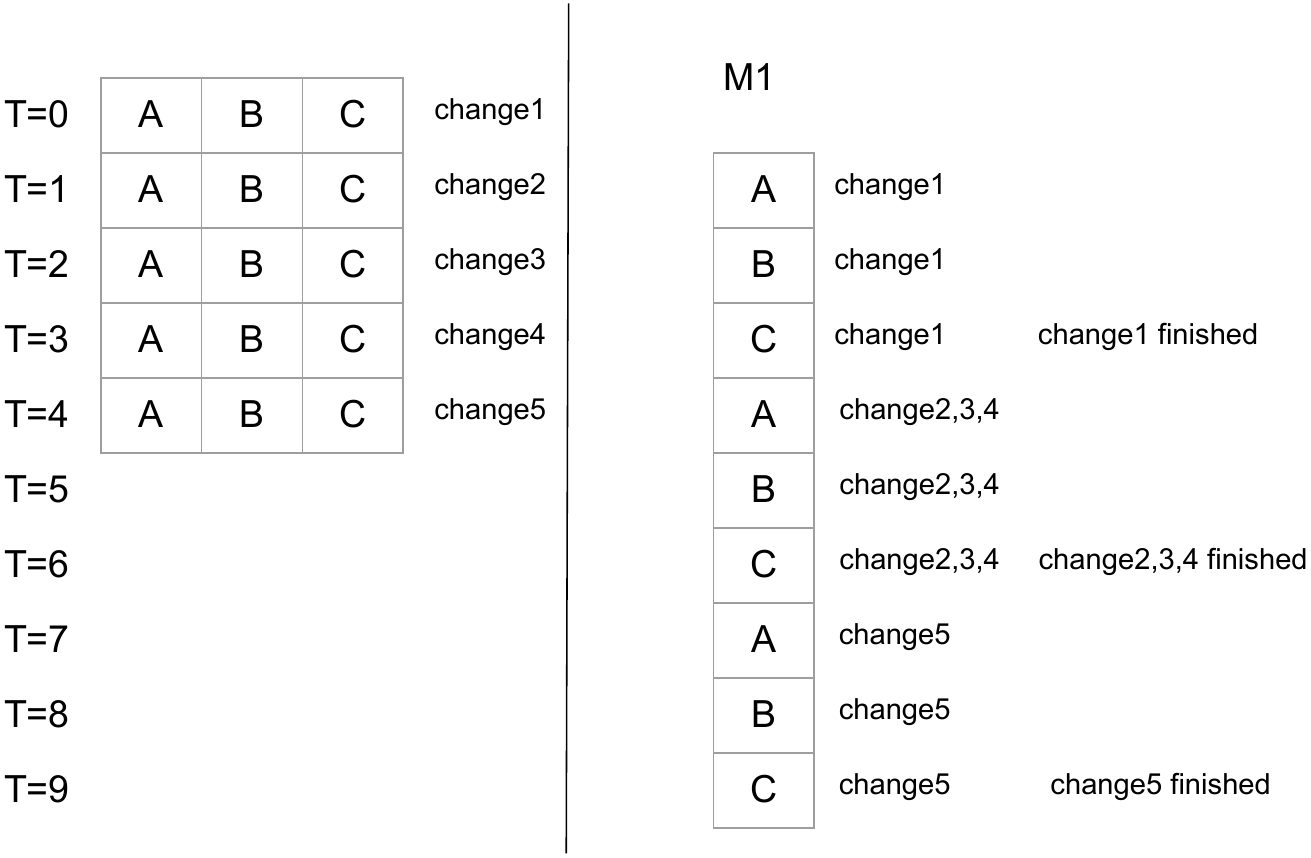}
	\caption[An example of \DynamicBatching methodology]{
  A sample of \DynamicBatching algorithm.
In this example, there are five subsequent changes and their requested tests A, B, and C.
The sequence of events is displayed by T, and the machines are shown by M. 
 }
	\label{fig:dynamic_batching}
\end{figure}

Figure \ref{fig:dynamic_batching} shows an example of \DynamicBatching with a single \mr. We assume that there is no failure and all the changes pass the tests. Using \DynamicBatching, after committing the first change, the resources are immediately allocated to it for testing. Change 1 arrives first, and only after it finishes testing Change 1, it batches all the changes that are now waiting, \ie changes 2, 3, and 4. After testing the second batch of size 3, it runs the tests for Change 5 in a batch of size 1 because no other changes are waiting for testing. 
The feedback time for each change will be 3, 5, 4, 3, and 5 respectively.

\subsection{TestCaseBatching}
\label{backgroundTCB}

\DynamicBatching can decrease the feedback time by reducing the idle time of resources. However, when all resources are utilized for testing, new changes must be queued until {\it all the tests} for the current batch are complete. 
With \TestCaseBatching new changes are added to the batch when any test finishes rather than having to wait for all the tests to finish. This approach requires the requested tests to be queued. To manage the test queue, the requested test cases for each change are added to the queue, \ie the ChangeID, and TestID. When a test finishes, any new changes are added to the batch and the next test in the queue is run. Once a change has had all its tests run, the results are reported. 

\begin{figure}
	\centering
	\includegraphics[width=\columnwidth]{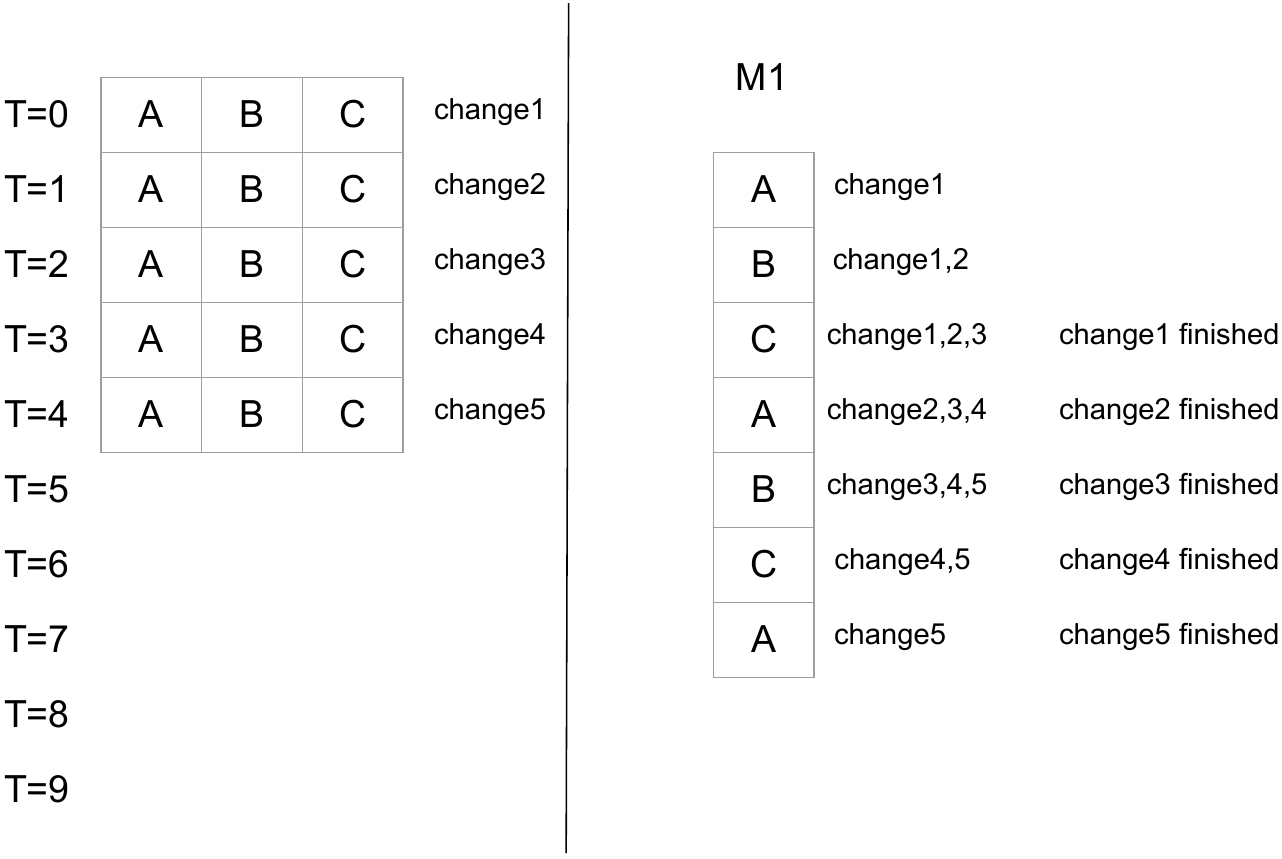}
	\caption[An example of TestCaseBatching]{
A sample of \TestCaseBatching algorithm.
In this example, there are five subsequent changes and their requested tests A, B, and C.
The sequence of events is displayed by T, and the machines are shown by M. 
	}
	\label{fig:testCase_batching}
\end{figure}

In Figure \ref{fig:testCase_batching} we provide an example of \TestCaseBatching.
After each test finishes, \TestCaseBatching includes any waiting builds and runs the next requested test in the test queue. \TestCaseBatching has to run Test A three times because it has finished for Change 1 before Changes 2, 3, and 4 and the algorithm has to run it independently for Change 5. In contrast, \TestCaseBatching must only run B and C twice as they overlap when more changes are available. 
We see that
the average feedback time is reduced to 3 compared to the 4 needed for \DynamicBatching, meaning that we get feedback to developers 25\% sooner.

\section{Results}
\label{sec_results}

This section presents the results of our evaluation, comparing the performance of different batching algorithms in terms of feedback time, the number of machines utilized, and the extent of execution reduction achieved. We conduct these evaluations using the \Ericsson and Chrome datasets, considering various numbers of machines.

\subsection{RQ1: Parallelization with \TestAll}
\begin{figure}
	\centering
	\includegraphics[width=\columnwidth]{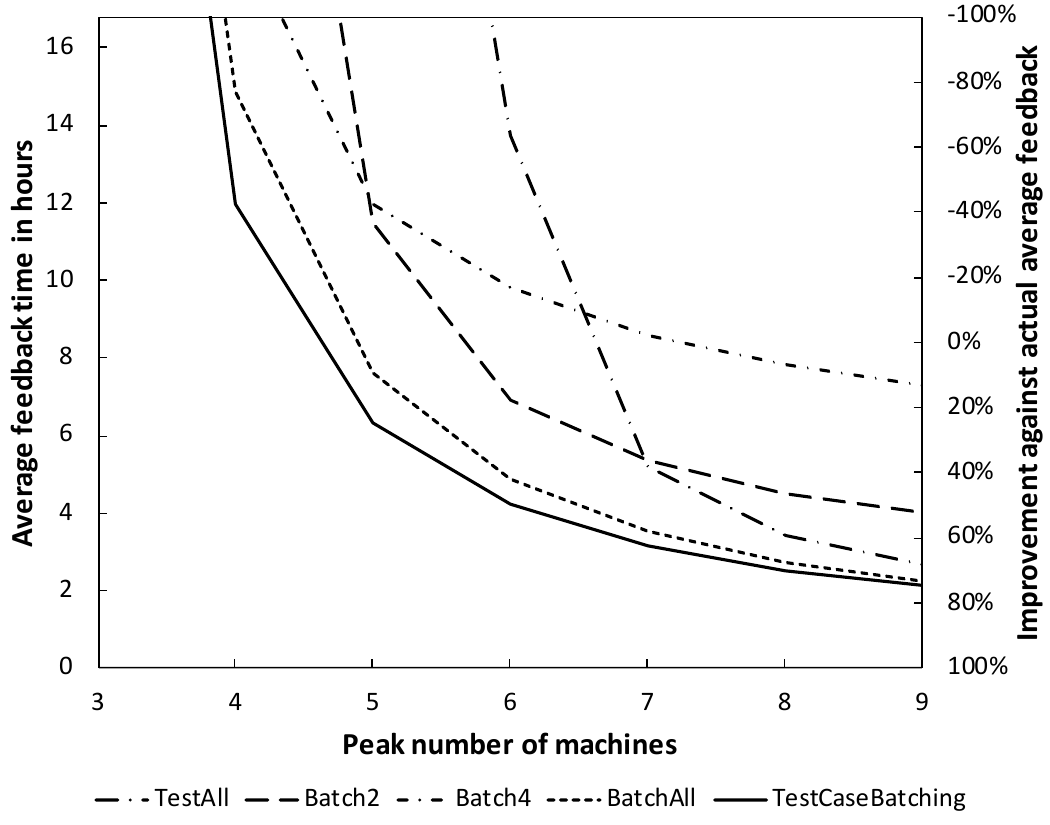}
         \caption{Average feedback time and percentage change relative to actual average feedback time for each approach with varying numbers of machines at \Ericsson.
         }
	\label{figEricssonFeedback}
\end{figure}

\begin{figure}
	\centering
	\includegraphics[width=\columnwidth]{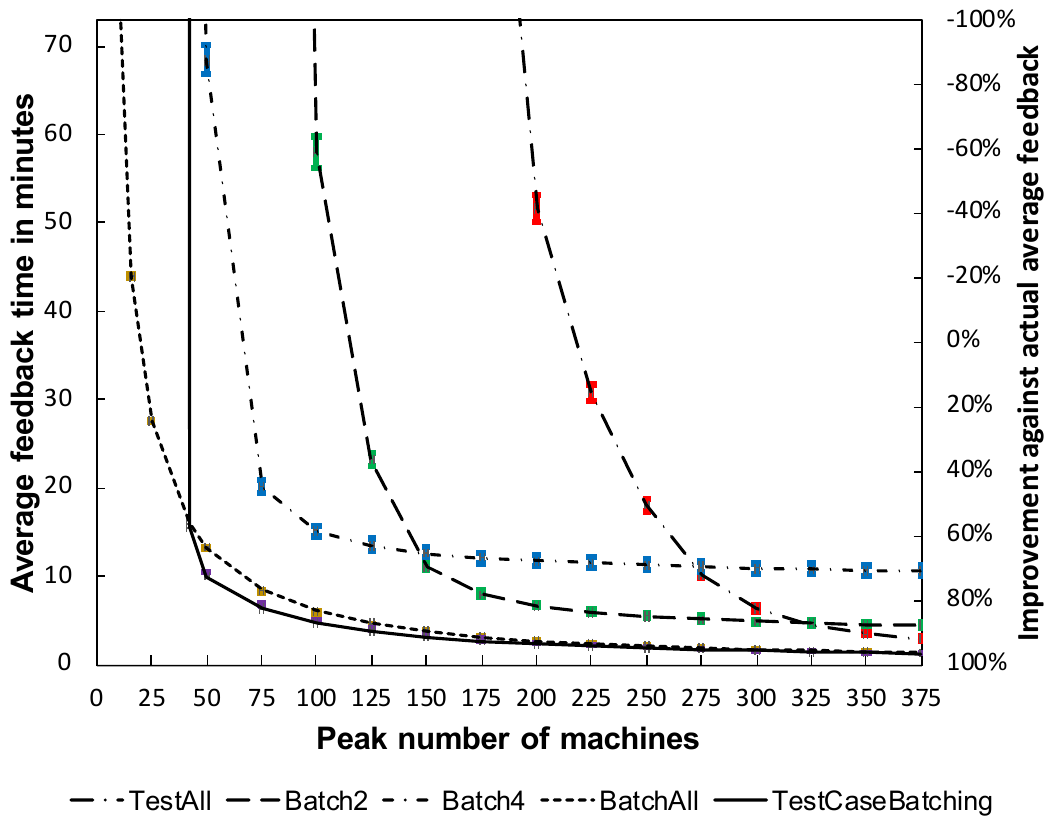}
         \caption{Average feedback time with colorful 95\% confidence intervals and percentage change relative to actual average feedback time for each approach with varying numbers of machines in Chrome.
         }
	\label{figChromeFeedback}
\end{figure}

In the \TestAll approach, we simulate testing each change individually to understand the impact of parallelization on the testing process.
The curves in Figure~\ref{figEricssonFeedback} and \ref{figChromeFeedback} show that
increasing the number of machines has a nonlinear impact on feedback time.
We see that \TestAll needs 7 \mrs to achieve the actual average feedback time of 8.33 hours for \Ericsson. For Chrome, we see \TestAll needs 217 available \mrs to achieve the 36.48 minutes average feedback time that is actually observed on Chrome. 

Substantial improvement is seen by adding additional parallel \mrs, but the return on investment diminishes at some point.
For instance, in \Ericsson, when we increase from 6 to 7 \mrs there is a speedup of 2.65 times in \TestAll algorithm, but the increase from 7 to 8 \mrs only adds 1.5 times of improvement in feedback time. In Chrome, \TestAll algorithm runs 49 times faster when we use 200 \mrs instead of 100 \mrs, while when we increase it from 200 to 300 \mrs it only gets 8.11 times faster. We see that \TestAll with 9 \mrs has plateaued for \Ericsson, and we see \TestAll with 350 \mrs has plateaued for Chrome.

Figure \ref{figChromeFeedback} displays the 95\% confidence intervals for the average feedback time values of all the algorithms analyzed in Chrome.\footnote{We are unable to add confidence intervals or compute statistical tests because we no longer have access to the \Ericsson data.}For the \TestAll algorithm, the confidence interval range becomes narrower, decreasing from 4\% to less than 1\% of the actual average feedback time, as the number of machines increases from 200 to 275 and beyond. This reduction in the width of the confidence interval indicates a decrease in the range of the feedback time distribution.

To compare the feedback time distributions of different batching algorithms for Chrome, we utilize the Wilcoxon Rank-Sum test, which is suitable for comparing two independent samples without relying on distribution assumptions. To address the issue of multiple comparisons, we apply the Bonferroni correction by dividing the desired overall p-value of 0.05 by the number of comparisons. This adjustment results in a p-value cutoff of 0.0003 for each comparison, ensuring a stringent threshold for statistical significance.

The obtained p-values are all significantly lower than the cutoff value, indicating significant differences in the feedback time distributions between most algorithm comparisons. However, the comparison between \TestAll and \BatchTwo at 300 machines yields a p-value of 0.06, suggesting a lack of significant difference in only this case.

We also calculate Cliff's delta effect size to measure the magnitude and direction of differences between feedback time distributions. In Table \ref{tableChromeEffectSize}, we present the effect sizes between various algorithms at different machine counts. This data further supports the observed trend and the significance of the differences depicted in the average feedback time plot shown in Figure \ref{figChromeFeedback}. 

\subsection{RQ2: \ConstantBatching}

Batching reduces the number of test executions when the changes request the same tests. Prior works~\cite{najafi_bisecting_2019, beheshtian_software_2021} used a constant batch size, and we replicate the state-of-the-art approach on a project at \Ericsson. We report the results for batch sizes 2 and 4 in this paper.

When large batches pass, common tests requested by changes are only run one time, dramatically reducing the amount of testing. As noted by prior work~\cite{beheshtian_software_2021}, the reduction is limited by the number of failing tests because a failure requires additional executions to find the culprit changes. However, on most large projects there are relatively few test failures, making batching highly effective \cite{fallahzadeh_impact_2022}.

Prior works have focused on resource savings and largely ignored or simplified feedback time~\cite{najafi_bisecting_2019, beheshtian_software_2021}. If we strictly follow a constant batch size, then we see that commits can wait for extended periods of time. 
Figure \ref{figEricssonFeedback} and \ref{figChromeFeedback} clearly reveal that constant batching algorithms, \BatchTwo and \BatchFour, outperform \TestAll in the majority of cases, except in highly resource-available environments where the number of machines significantly exceeds the baseline.
As the number of resources increase, they plateau with a relatively high feedback time and \TestAll outperforms them. However, they are simpler to implement than other batching algorithms.

Table \ref{tableMachinesRequiredToHoldBaselineAverageFeedbackTime} shows that \BatchTwo effectively maintains the actual average feedback baseline of 8.33 hours for \Ericsson using 6 \mrs, resulting in a 14.29\% reduction in machine usage compared to the baseline. However, \BatchFour achieves the same average feedback time using 8 machines, resulting in a negative reduction (-14.29\%) in machine usage compared to the baseline. 
For Chrome, \BatchTwo can maintain the actual average feedback baseline of 36.48 minutes using 113 machines, resulting in a 47.93\% reduction in machine usage compared to the baseline. Similarly, \BatchFour achieves the same average feedback time using 61 machines, resulting in a 71.89\% reduction in machine usage compared to the baseline.

Table \ref{tableEricssonRequiredMachinesToPlateau} shows that \BatchTwo and \BatchFour plateau for \Ericsson at 4.52 and 7.86 hours by using 8 \mrs respectively.
This means that \BatchTwo and \BatchFour reach a plateau with 70\% and 195\% longer average feedback time compared to the baseline, respectively. Consequently, increasing the number of machines for both algorithms provides a limited scope for improvement compared to the baseline.
For Chrome, Table \ref{tableChromeRequiredMachinesToPlateau} displays that \BatchTwo and \BatchFour plateau at 5.88 and 12.6 minutes using 225 and 150 \mrs. 
This results in 31\% and 255.93\% longer plateaued feedback time compared to the baseline, respectively, indicating limited room for improvement relative to the baseline.

The confidence interval values displayed in Figure \ref{figChromeFeedback} for the \BatchTwo algorithm and Chrome reveal that the average feedback time has a range of around 5\% when utilizing 100 machines. Notably, as the number of machines increases to 150, this range significantly decreases to less than 1\%. Conversely, for the \BatchFour algorithm, the confidence interval values indicate a range of about 5\% when using 50 machines. While increasing the number of machines reduces the range, this reduction is constrained, and even with 375 machines, the range remains at 2\%.

Figures \ref{figEricssonExecutionReduction} and \ref{figChromeExecutionReduction} illustrate the percentage of \ExecutionReduction achieved by various batching algorithms on \Ericsson and Chrome, respectively. It is worth noting that the \ExecutionReduction for the constant batching algorithms remains constant irrespective of the number of machines used. For \Ericsson, \BatchTwo achieves an \ExecutionReduction of 23\%, while for Chrome, it achieves an \ExecutionReduction of 50\%. Similarly, \BatchFour achieves a \ExecutionReduction of 37\% for \Ericsson and 75\% for Chrome.

\begin{table}
\centering
\caption{Number of machines required to maintain average feedback time in \textbf{\Ericsson} and \textbf{Chrome}, along with percentage reduction compared to baseline.}
\resizebox{\columnwidth}{!}{%
\setlength\tabcolsep{2.5pt} 
\begin{tabular}{@{}llllll@{}}
\toprule
\textbf{Algorithm} & \textbf{\TestAll} & \textbf{\BatchTwo} & \textbf{\BatchFour} & \textbf{\DynamicBatching} & \textbf{\begin{tabular}[c]{@{}l@{}}\textit{TestCase}\\\textit{Batching}\end{tabular}} \\
\midrule
\textbf{\Ericsson} & 7 (0\%) & 6 (14.29\%) & 8 (-14.29\%) & 5 (28.57\%) & 5 (28.57\%) \\
\textbf{Chrome} & 217 (0\%) & 113 (47.93\%) & 61 (71.89\%) & 20 (90.78\%) & 42 (80.65\%) \\
\bottomrule
\end{tabular}%
}
\label{tableMachinesRequiredToHoldBaselineAverageFeedbackTime}
\end{table}

\begin{table}
\centering
\caption{The number of \mrs and the average feedback time in hours where each batching algorithm plateaus for \Ericsson.}
\resizebox{\columnwidth}{!}{%
\setlength\tabcolsep{2.5pt} 
\begin{tabular}{@{}llllll@{}}
\toprule
\textbf{Algorithm} & \textbf{\begin{tabular}[c]{@{}l@{}}\textit{TestCase}\\\textit{Batching}\end{tabular}} & \textbf{\DynamicBatching} & \textbf{\TestAll} & \textbf{\BatchTwo} & \textbf{\BatchFour} \\ 
\midrule
\textbf{Feedback time} & 2.53           & 2.75                      & 2.66              & 4.52               &  7.86  \\
\textbf{Machines}    & 8              & 8                         & 9                 & 8                  & 8   \\
\bottomrule
\end{tabular}
}
\label{tableEricssonRequiredMachinesToPlateau}
\end{table}

\begin{table}
\centering
\caption{The number of \mrs and the average feedback time in minutes where each batching algorithm plateaus for Chrome.}
\resizebox{\columnwidth}{!}{%
\setlength\tabcolsep{2.5pt} 
\begin{tabular}{@{}llllll@{}}
\toprule
\textbf{Algorithm} & \textbf{\begin{tabular}[c]{@{}l@{}}\textit{TestCase}\\\textit{Batching}\end{tabular}} & \textbf{\DynamicBatching} & \textbf{\TestAll} & \textbf{\BatchTwo} & \textbf{\BatchFour} \\
\midrule
\textbf{Feedback time} & 3.96                       & 3.78                     & 3.54             & 5.88            & 12.6            \\
\textbf{Machines}    & 125                       & 150                      & 350              & 225             & 150             \\
\bottomrule
\end{tabular}
}
\label{tableChromeRequiredMachinesToPlateau}
\end{table}

\begin{table*}
\centering
\caption{The Cliff's delta effect size values for pairwise comparisons of different batching algorithms for Chrome at different numbers of machines.}
\resizebox{\textwidth}{!}{%
\setlength\tabcolsep{4pt} 
\begin{tabular}{@{}lllllllllll@{}}
\toprule
\textbf{Machines} & \textbf{TA vs B2} & \textbf{TA vs B4} & \textbf{TA vs BA} & \textbf{TA vs TCB} & \textbf{B2 vs B4} & \textbf{B2 vs BA} & \textbf{B2 vs TCB} & \textbf{B4 vs BA} & \textbf{B4 vs TCB} & \textbf{BA vs TCB} \\
\midrule
8 & 0.53 & 0.79 & 1.00 & -0.05 & 0.57 & 0.99 & -0.30 & 0.98 & -0.51 & -0.92 \\
16 & 0.57 & 0.84 & 0.99 & -0.04 & 0.68 & 0.98 & -0.30 & 0.97 & -0.53 & -0.87 \\
25 & 0.63 & 0.90 & 0.98 & 0.08 & 0.83 & 0.97 & -0.17 & 0.94 & -0.48 & -0.67 \\
50 & 0.83 & 0.94 & 0.96 & 0.97 & 0.84 & 0.93 & 0.95 & 0.65 & 0.80 & 0.51 \\
75 & 0.90 & 0.94 & 0.95 & 0.96 & 0.62 & 0.78 & 0.85 & 0.52 & 0.73 & 0.43 \\
100 & 0.83 & 0.87 & 0.91 & 0.92 & 0.33 & 0.63 & 0.75 & 0.50 & 0.70 & 0.35 \\
125 & 0.65 & 0.69 & 0.80 & 0.84 & 0.12 & 0.55 & 0.67 & 0.54 & 0.70 & 0.29 \\
150 & 0.60 & 0.58 & 0.74 & 0.78 & -0.08 & 0.49 & 0.62 & 0.58 & 0.70 & 0.24 \\
175 & 0.48 & 0.40 & 0.66 & 0.71 & -0.19 & 0.46 & 0.59 & 0.61 & 0.71 & 0.20 \\
200 & 0.34 & 0.20 & 0.59 & 0.64 & -0.25 & 0.45 & 0.56 & 0.63 & 0.71 & 0.16 \\
225 & 0.24 & 0.06 & 0.52 & 0.58 & -0.28 & 0.45 & 0.55 & 0.65 & 0.72 & 0.14 \\
250 & 0.16 & -0.06 & 0.47 & 0.53 & -0.30 & 0.45 & 0.54 & 0.67 & 0.73 & 0.12 \\
275 & 0.06 & -0.20 & 0.42 & 0.49 & -0.32 & 0.46 & 0.53 & 0.68 & 0.73 & 0.11 \\
300 & -0.02 & -0.31 & 0.39 & 0.45 & -0.33 & 0.46 & 0.53 & 0.69 & 0.73 & 0.10 \\
325 & -0.09 & -0.39 & 0.35 & 0.42 & -0.34 & 0.46 & 0.52 & 0.70 & 0.74 & 0.09 \\
350 & -0.14 & -0.45 & 0.33 & 0.39 & -0.34 & 0.47 & 0.52 & 0.70 & 0.74 & 0.08 \\
375 & -0.18 & -0.49 & 0.30 & 0.36 & -0.35 & 0.47 & 0.52 & 0.71 & 0.74 & 0.08 \\
\bottomrule
\end{tabular}%
}
\label{tableChromeEffectSize}
\small
$\textsc{*}$Please refer to the following abbreviations used in this table:
TA: \TestAll,
B2: \BatchTwo,
B4: \BatchFour,
BA: \DynamicBatching,
TCB: \TestCaseBatching.

\end{table*}


\subsection{RQ3: \DynamicBatching}

\ConstantBatching excels in conserving resources and reducing feedback time in highly resource-constrained environments characterized by limited machine availability compared to the baseline and a significant backlog of commits. However, the queue size varies over time, with peak changes happening during working hours. To better utilize the available resources, we suggest \DynamicBatching which batches all available changes in the queue.

Figure \ref{figEricssonFeedback} and Figure \ref{figChromeFeedback} show that \DynamicBatching approach always outperforms \TestAll and \ConstantBatching algorithms in both \Ericsson and Chrome cases. This algorithm performs promisingly both in high resource-constrained and high resource-available conditions compared to the baseline number of machines.


Table \ref{tableMachinesRequiredToHoldBaselineAverageFeedbackTime} illustrates that \DynamicBatching is capable of maintaining the feedback baseline at 8.33 hours for \Ericsson by utilizing 5 machines, leading to a 28.57\% reduction in machine usage. Similarly, for Chrome, \DynamicBatching achieves the feedback baseline at 36.48 minutes with 20 machines, resulting in an impressive 90.78\% reduction in machine utilization.

Table \ref{tableEricssonRequiredMachinesToPlateau} shows that \DynamicBatching approach plateaus for \Ericsson at 2.75 hours by using 8 \mrs. This represents a 12.5\% resource reduction relative to the plateaued feedback baseline. For Chrome, Table \ref{tableChromeRequiredMachinesToPlateau} displays that \DynamicBatching plateaus for Chrome at 3.78 minutes using 150 \mrs. This is a 54.14\% reduction in resources to reach the plateaued feedback baseline.

The confidence interval for the average feedback time values depicted in Figure \ref{figChromeFeedback} for the \DynamicBatching algorithm and Chrome indicates a range of approximately 1\% compared to the actual average feedback time when using 16 machines. As the number of machines increases to 25, this range narrows significantly, approaching zero, which suggests a more concentrated distribution of the feedback time values.

\begin{figure}
	\centering
	\includegraphics[width=1\columnwidth]{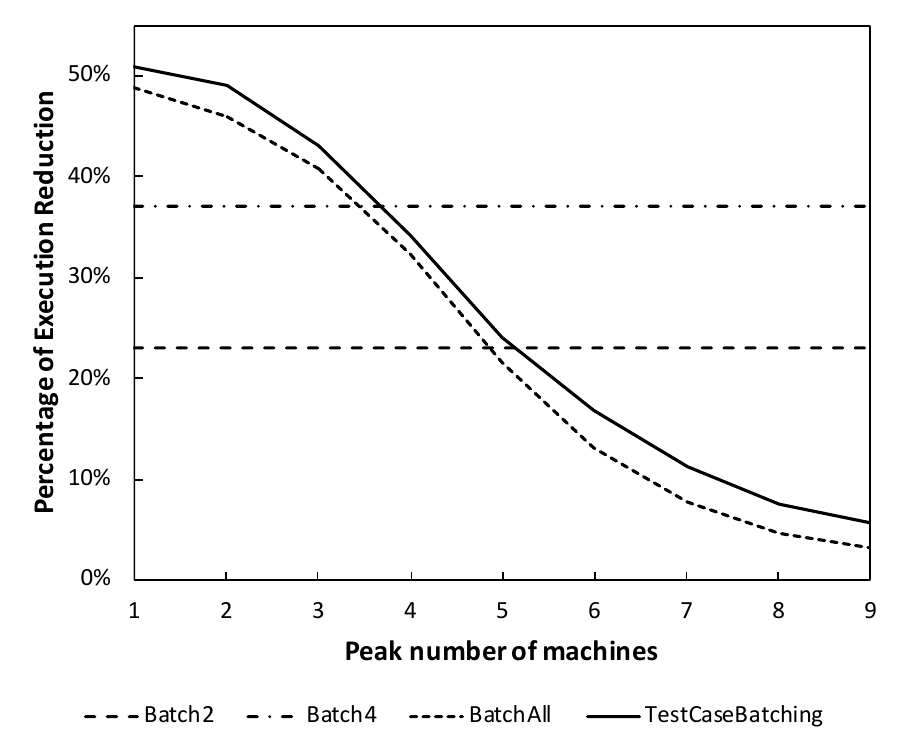}
        \caption{The percentage of \ExecutionReduction achieved by different batching algorithms, relative to the \TestAll, was calculated for different peak numbers of machines for \Ericsson.}
	\label{figEricssonExecutionReduction}
\end{figure}

\begin{figure}
	\centering
	\includegraphics[width=1\columnwidth]{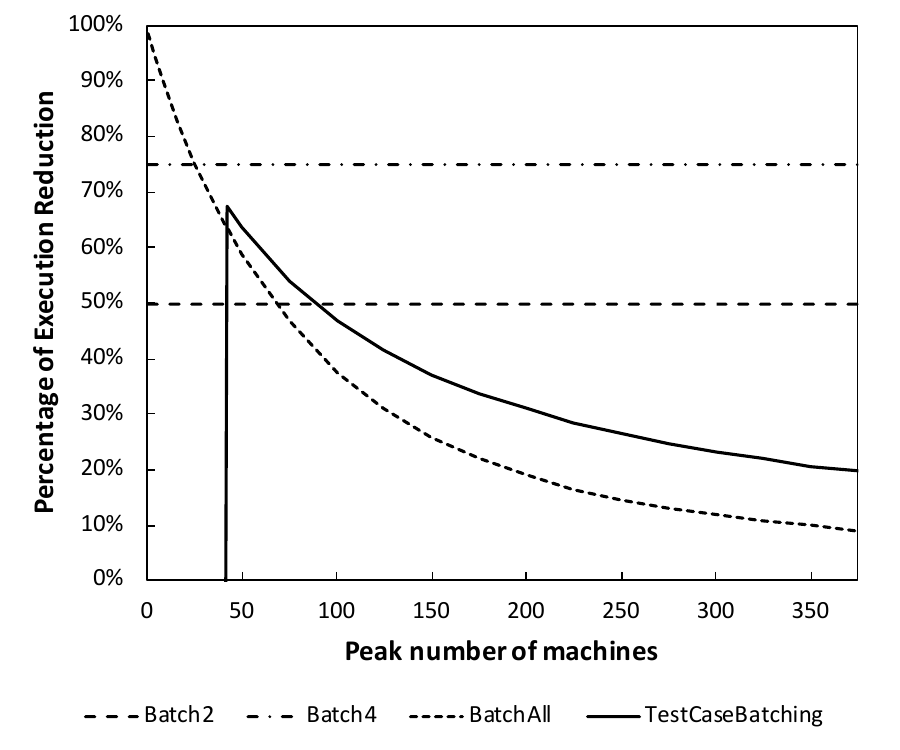}
	\caption{The percentage of \ExecutionReduction achieved by different batching algorithms, relative to the \TestAll, was calculated for different peak numbers of machines for Chrome.}
	\label{figChromeExecutionReduction}
\end{figure}

Figure \ref{figEricssonExecutionReduction} and \ref{figChromeExecutionReduction} demonstrate that the \DynamicBatching algorithm's \ExecutionReduction varies with the number of machines used, decreasing as the number of machines increases. For \Ericsson, the \ExecutionReduction ranges from 49\% to 3\% when utilizing 1 to 9 machines. Similarly, for Chrome, the \ExecutionReduction ranges from 99\% to 9\% when employing 1 to 375 machines.  

Figure \ref{figChromeAverageBatchSizes} depicts the average batch sizes utilized by the \DynamicBatching and \TestCaseBatching algorithms in Chrome, with varying numbers of machines. The figure is displayed on a logarithmic scale to accommodate the substantial difference in batch sizes between a few machines and larger numbers of machines. The average batch sizes for the \DynamicBatching algorithm range from approximately 85 batches to nearly 1 batch as the number of machines increases from 1 to 375. Notably, a significant reduction in average batch sizes is observed when the number of machines increases from 1 to 8, resulting in a drop from 85 to 10 batches for \DynamicBatching.

\subsection{RQ4: \TestCaseBatching}

\DynamicBatching processes all available changes. However, any change that arrives has to wait until all the tests for a batch have been completed. In background Section~\ref{backgroundTCB}, we introduced \TestCaseBatching that queues the requested tests across all changes and includes any new change after each test completes (rather than waiting for all tests to complete).

Except when there is an extreme resource constraint, the \TestCaseBatching approach performs more effectively than other algorithms in terms of feedback time as shown in for \Ericsson in Figure \ref{figEricssonFeedback} and Chrome in Figure \ref{figChromeFeedback}.
The reasons for the algorithm's poor performance under high resource constraints, as depicted by the parallel line in Figure \ref{figChromeFeedback} at 42 machines, will be discussed in detail in the discussion section.

Table \ref{tableMachinesRequiredToHoldBaselineAverageFeedbackTime} presents that \TestCaseBatching achieves the feedback baseline at 8.33 hours for \Ericsson by utilizing 5 machines, resulting in a 28.57\% reduction in machine usage. Similarly, for Chrome, \TestCaseBatching achieves the feedback baseline at 36.48 minutes using 42 machines, leading to a remarkable 80.65\% reduction in machine utilization.

Table \ref{tableEricssonRequiredMachinesToPlateau} shows that the \TestCaseBatching approach plateaus for \Ericsson at 2.53 hours by using 8 \mrs. This presents a 12.5\% resource reduction to reach the plateaued feedback baseline. For Chrome, Table \ref{tableChromeRequiredMachinesToPlateau} displays that the \TestCaseBatching plateaus for Chrome at 3.96 minutes using 125 \mrs. This is a 61.28\% reduction in resources to reach the plateaued feedback baseline.

The confidence interval ranges depicted in Figure \ref{figChromeFeedback} for the \TestCaseBatching algorithm and Chrome tend to stabilize at approximately 1\% of the actual average feedback time as the number of machines increases to 50. Beyond this threshold, the range of confidence intervals decreases significantly, approaching nearly zero. This narrowing range suggests a more concentrated distribution of feedback time values.

Figures \ref{figEricssonExecutionReduction} and \ref{figChromeExecutionReduction} illustrate the varying percentages of \ExecutionReduction achieved by the \TestCaseBatching algorithm using different numbers of machines. In the context of \Ericsson, \TestCaseBatching achieves an \ExecutionReduction ranging from 51\% to 6\% when employing 1 to 9 machines. On the other hand, in the case of Chrome, \TestCaseBatching initially exhibits negative performance in \ExecutionReduction with fewer than 42 machines, as further discussed in the subsequent section. However, with 42 to 375 machines, \TestCaseBatching achieves an \ExecutionReduction ranging from 67\% to 20\%, outperforming the results of \DynamicBatching.

\begin{figure}
	\centering
	\includegraphics[width=\columnwidth]{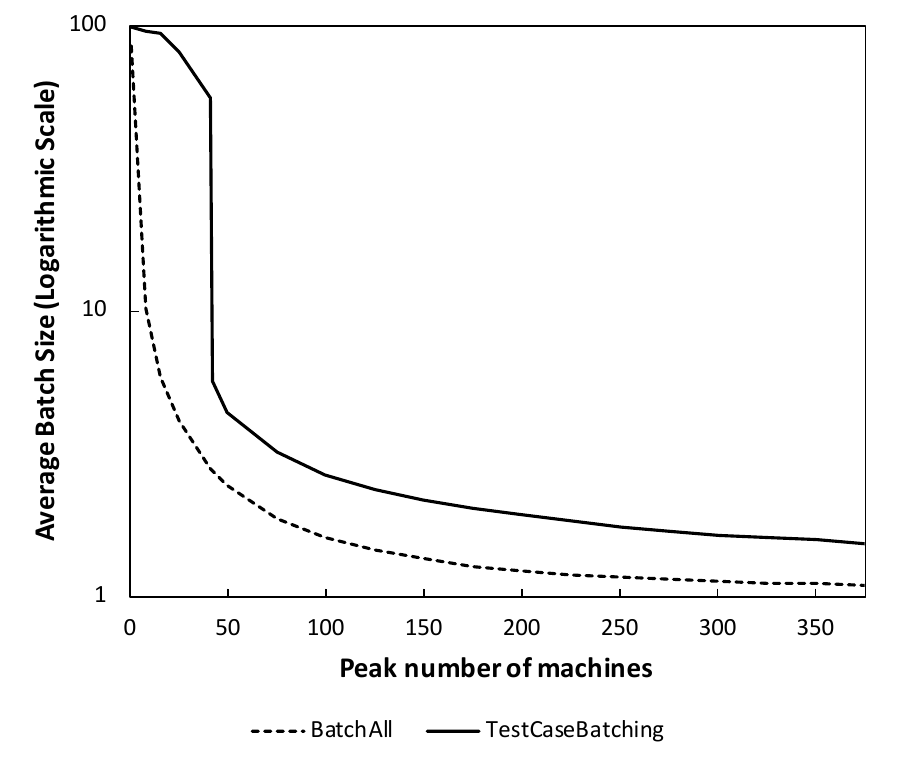}
	\caption{The average batch size used by the \DynamicBatching and \TestCaseBatching algorithms was calculated for different peak numbers of machines in Chrome, represented on a logarithmic scale.}
	\label{figChromeAverageBatchSizes}
\end{figure}

Figure \ref{figChromeAverageBatchSizes} illustrates the batch size utilization of the \TestCaseBatching algorithm on Chrome, covering a range of 1 to 375 machines. The average batch sizes vary from approximately 95 to 1.5. In the range of 1 to 41 machines, the average batch size remains relatively high, ranging from 98 to 56. However, at 42 machines, a significant drop occurs, resulting in an average batch size of around 6. Beyond this point, the number of average batch sizes continues to decrease gradually, reaching 1.5 at 375 machines.

\section{Discussion}
\label{sec_discussion}


Regarding the impact of parallelization on testing with resource constraints, an increase in machine count non-linearly improves feedback time. Initially, limited resources lead to significant delays, as subsequent changes compound the delay. For instance, with just one \mr, the queue grows substantially, causing wait times to surpass processing times. As the queue size decreases, improvements become smaller, with test execution time becoming the key factor. Eventually, feedback time plateaus, offering negligible gains from further machine increases due to the small queue size.

The results for the \ConstantBatching algorithms of \BatchTwo and \BatchFour reveal that running these algorithms in a highly resource-constrained environment can boost their feedback time in comparison with the \TestAll approach. Moreover, the more resource-constrained the environment is, the more effective larger batch sizes would be. 
However, in a highly resource-available situation, using a larger constant batch size will deteriorate the feedback time. This is because of the fewer changes remaining in the queue as a result of the faster processing time of the changes. Consequently, \ConstantBatching with large batch sizes becomes a disadvantage since it increases the wait time for processing the changes. 
This is also the reason why \ConstantBatching algorithms reach a plateau at higher feedback time values, leaving little room for improvement with the utilization of additional machines.

\DynamicBatching on the other hand uses the fast unloading advantage of a bigger batch size in a high resource-constrained environment and the low wait time benefit of a smaller batch size in a high resource-available environment. 
When we simulate a high resource-constrained environment with few \mrs, it makes large batches that can rapidly reduce the queue, leading to a higher \ExecutionReduction. Conversely, being in a high resource-available condition with enormous \mrs, it may reduce the batch size to 1, eliminating build waiting time but resulting in a lower \ExecutionReduction.
This makes \DynamicBatching highly effective in terms of feedback time in both of these scenarios.

\TestCaseBatching algorithm has the highest performance in terms of feedback time, except in an extremely high resource-constrained environment with few \mrs, specifically in the Chrome case with fewer than 42 machines. 
This is because the \DynamicBatching algorithm puts some changes in the queue while it is running the current batch, and then it runs all its tests as a batch. Still, the \TestCaseBatching method always accepts new changes and does not adopt a queue. Since it runs the tests of a change in the absence of potential subsequent changes, when the next changes become available, it has to run some of these tests again as a penalty. When there are enormous overlapping builds, these penalties add up and deteriorate the feedback time and the \ExecutionReduction of the \TestCaseBatching algorithm in comparison with the \DynamicBatching method. This specifically is the case in Chrome test results, as there are more overlapping builds in the dataset.

The implications of these results for practitioners would be as follows.
We recommend using \DynamicBatching and \TestCaseBatching as the best batching algorithms among all the approaches under the study.
In a very high resource-constrained environment, \DynamicBatching produces a better feedback time and \ExecutionReduction, otherwise, \TestCaseBatching performs the best.
Despite the testing approach being used, there is a threshold in which increasing the number of machines does not improve the feedback significantly.
Hence, it makes sense to recognize this plateau point by exercising the batching approach being used on the corresponding project.
This way they can reduce feedback time while saving the number of machines being used. 


\section{Threats to validity}
\label{sec_threats}


\textbf{\textit{External Validity.}}
%
%
The outcomes from this study are from applying various batching techniques on the test results of two real large-scale \Ericsson and Chrome projects, and we may not be able to generalize them to all other applications.
The performance of these batching algorithms might differ by applying them to various other projects and using different numbers of machines. 
This is because they probably have different numbers of changes, builds, and test cases.
However, the algorithms adopted in this study are not tied to a specific project, and the non-linear relationship between the number of machines and the feedback time in other projects would likely be the same.

Another issue that may threaten the external validity of this work is the failure rate.
The failure rate can be different among various applications, and the higher the failure rate is, the higher the cost of culprit finding would become which in turn, can make the batching algorithms ineffective.
However, Beheshtian \etal \cite{beheshtian_software_2021} show that the batching algorithms can produce savings on the projects with a build failure ratio of below 40\%, and among 9 Travis CI projects under their study, 85.5\% of them could take advantage of batching techniques.
The build failure rate in Chrome dataset is 8.5\%, which is well below the 40\% threshold, making batching an effective approach for this project.
Meanwhile, the focus of this study is more on the impact of parallelism on various batching algorithms and on two very large-scale \Ericsson and Chrome projects. 
At the time of this study, we could not recognize any other available testing datasets for the experiment which are comparable in testing scale with these datasets.

\textbf{\textit{Internal Validity.}}
%
One of the aspects of test optimization which can compromise the process is test dependencies. Reordering or running tests in parallel when they were not designed to be run in this manner can introduce dependency flaky failures~\cite{Lam2019ICST}.
Therefore, we only batch tests that \Ericsson guarantees are independent.
In the case of Chrome, we only run tests in parallel and independently if Chrome developers were already running these in parallel on multiple shards.

The historical simulation simplified parts of the \Ericsson's and Chrome's testing processes. 
For example, developers can stop testing a build or manually batch-select changes for testing. 
Since we cannot model these manual interventions, we exclude them from our simulation. 

In our experiment, we assumed that all changes can be batched and none lead to merge conflict. 
Since each must be ultimately merged into the main branch, we do not introduce any new conflicts because any conflict would have been dealt with when the developer ensures that the code can be merged. 
However, the batching process may bring this conflict to the developer's attention earlier as we batch different combinations of changes.

\textbf{\textit{Construct Validity.}}
To simplify the measurement of the feedback time we ignored the compile times of the changes.
This is because we could not have any assumptions for the compile time of the batches.
In the case that the batch contains no failing change, this can save on compile time.
Otherwise, it might cause some extra compile time for the failing changes.
Considering the failure rates in our study, we can suggest that calculating the compile time can even save more on the feedback time of the batching techniques. We leave this simulation parameter to future work.

Using the average value to compare the feedback time of the different batching algorithms can be a threat to the construct validity of this study.
Although the feedback time distribution of each test batching experiment is a more accurate way of showing the results, for each batching algorithm and for each number of machines, we need a single value to compare the performance.
This along with having an approximately normal distribution, lead us to determine the average feedback time to compare the outcomes.


\section{Related Work}
\label{sec_related_works}

Continuous integration and delivery (CI/CD) is a crucial part of modern software development \cite{jin2021helped}. It enables developers to automatically build and test changes, ensuring software functionality remains intact \cite{poth_how_2018, hilton_understanding_2016, soni_end_2015, leppanen2015highways}. However, testing every commit individually in large software systems is often impractical \cite{herzig_art_2015, elbaum_techniques_2014}. In the following, we investigate approaches to accelerate CI.

\textbf{\textit{Test Selection and Prioritization.}} 
To reduce both resource consumption and provide earlier feedback, test selection has been widely adopted in the industry and extensively studied ~\cite{536955, ENGSTROM201014, chen_using_2011, machalica_predictive_2019, parthasarathy_rtl_2022, arrieta_multi-objective_2022, elsner_build_2022, verma_novel_2023, birchler_cost-effective_2023}. However, test selection's inherent trade-off is that not all tests are run, and some test failures may be missed. 
In contrast, test prioritization guarantees that all tests will be run, but those that are more likely to reveal faults will be run first, reducing feedback time on test failures, but not reducing the resource consumption in testing ~\cite{bagherzadeh_reinforcement_2022,  zhu_test_2018,962562, jahan_risk-based_2020, sharif_deeporder_2021, huang_dissimilarity-based_2022, mahdieh_test_2022,nayak_analytic_2022, bajaj_improved_2022, yaraghi_scalable_2023}. 

However, batching offers both feedback and execution reduction without missing any failures. To provide an industrial comparison, we contrast our results with those obtained at Microsoft. Herzig et al. \cite{herzig_art_2015} reported improvements of 40.31\%, 40.10\%, and 47.45\% in testing feedback time for Windows, Office, and Dynamics using association rule mining for selection. However, they noted slip-through rates of 71\% and 91\% at the first branch level, 21\% and 3\% at the second, and 8\% and 0\% at the third branch level. At Ericsson, the same approach resulted in a 42.78\% reduction but with 34.65\% slip-throughs \cite{najafi_improving_2019}. In contrast, \DynamicBatching and \TestCaseBatching reduced testing time by 57.44\% and 61.87\%,
respectively, at Ericsson without allowing any slip-throughs.

\textbf{\textit{Test parallelization.}}
Tests are distributed across machines using this technique to reduce feedback time. Previous works widely studied the impact of test parallelization on software testing and introduced algorithms to run tests in parallel~\cite{10.1145/581339.581397, 10.1145/1287624.1287645, bagies2020parallelizing, landing_cluster-based_2020}.
For example, Arabnejad \etal \cite{arabnejad_autopar-clava_2018} investigated using GPUs for running tests in parallel. 
The most popular algorithms for parallelizing tests are scheduling tests across the \mrs based on their IDs and their historical execution time \cite{shashban_run_2022}.

Candido et al. \cite{candido_test_2017} analyzed over 450 Java projects, discovering that less than 20\% of major projects utilize test parallelization due to concurrency concerns. They recommended strategies like test refactoring and grouping based on dependencies to aid parallel testing. Bell et al. \cite{10.1145/2786805.2786823} investigated test dependency's impact on parallelization, introducing the ElectricTest approach. Ding et al. \cite{ding2007software} proposed behavior-oriented test parallelization. Nagy et al. \cite{nagy_enhanced_2022} designed a parallel architecture for test case dispatching to reduce idle nodes and enhance execution time. In the context of \Ericsson and Chrome, we exclusively batch tests at levels designed for parallel execution. Notably, none of these prior studies explored test parallelization within batch testing. 

\textbf{\textit{Build Prediction and Skip.}}
Another line of work aimed at expediting Continuous Integration (CI) processes includes build prediction and build skip techniques. 
%
Build prediction techniques leverage machine learning methods to predict the result of a build, with a particular focus on reducing the cost of builds that are likely to pass \cite{finlay_data_2014,hassan_using_2006, hassan_change-aware_2017,kwan_does_2011,ni_cost-effective_2017,schroter_predicting_2010,wolf_predicting_2009,xie_cutting_2018,chen_buildfast_2020,saidani_prediction_2020}.
On the other hand, build skip techniques aim to identify builds that do not require execution, typically due to the absence of source code changes. There are different approaches for build skip, including manual configuration \cite{noauthor_travis_nodate,noauthor_using_nodate-1} and rule-based or learning-based methods \cite{abdalkareem_machine_2021,abdalkareem_which_2021,jin_cost-efficient_2020,jin_which_2022,jin_hybridcisave_2023,gallaba_accelerating_2022,saidani_detecting_2022}. 

While build prediction and build skip techniques offer potential benefits for accelerating CI processes, they have limitations. Build prediction techniques struggle with accurately predicting failing builds, leading to costly misidentifications, and their practical use cases are often ill-defined \cite{chen_buildfast_2020}. On the other hand, build skip techniques risk skipping builds that may contain relevant changes or introduce failures. In contrast, batching combines multiple changes into a single build, eliminating the risk of skipping builds entirely. It enables efficient debugging and failure resolution within the batch, providing a reliable approach for managing complex software changes in continuous integration, as employed by major companies like Google.


\textbf{\textit{Batch Testing.}}
This technique, employed in resource-constrained environments \cite{cho_adaptive_2017, chang_optimal_2009}, reduces feedback time. Instead of testing each change individually, changes are batched and tested simultaneously. GitBisection \cite{manpages2015git}, a well-known culprit-finding technique, conducts a binary search to identify the culprit, typically requiring $log(n)$ executions. However, when multiple culprits exist, GitBisection finds only the first. To address this, Najafi \etal \cite{najafi_bisecting_2019} introduced a bisection with a divide-and-conquer algorithm, capable of identifying all culprits. The total executions range from $2log(n)$ to $2n+1$ when all batch changes are culprits. Beheshtian \etal \cite{beheshtian_software_2021} showed that for batches of 4 or fewer, bisection increases executions. They propose \BatchStopFour, testing each change individually in batches of 4 or fewer. \BatchStopFour consistently outperforms batch bisection mathematically. Their study also examines failure rate effects on various batching algorithms.

Bavand \etal~\cite{bavand_mining_2021} implemented a complex dynamic batch size approach based on historical failure rates, resulting in a modest 5.17\% reduction in test execution time compared to \BatchFour with a single machine. In the \Ericsson context, \DynamicBatching and \TestCaseBatching achieved more significant \ExecutionReduction, with 11.87\% and 13.92\% reductions respectively compared to \BatchFour. For Chrome, \DynamicBatching showed a substantial 23.88\% improvement over \BatchFour with a single machine, while \TestCaseBatching had negative results with less than 42 machines. Overall, \DynamicBatching and \TestCaseBatching consistently outperformed \ConstantBatching in feedback time across different machine configurations.
These previous works focused on batch testing at the change level and with a single \mr, without exploring the effects of parallel testing. In our study, we investigate the combined impact of batching and parallel testing for the first time, examining how varying the number of parallel \mrs affects two extensive projects: \Ericsson and Chrome. 

\section{Conclusions and Future Work}
\label{sec_conclusion}
In this study, we aimed to investigate the impact of parallelization on different batching techniques, addressing the limitations of prior single-machines studies.
Our evaluation yielded the following findings. \TestAll experience compounded delays in subsequent builds, and the effect of changing the number of machines on feedback time is non-linear. The performance of \ConstantBatching is better in high resource-constrained setups, and it plateaus at a longer feedback time.
It provides a consistent \ExecutionReduction across different machine counts. 
\DynamicBatching is effective in both high resource-constrained and highly available resource environments, being able to maintain actual average feedback time and attain plateaued feedback baseline with 90.78\% and 54.14\% fewer machines than the baseline respectively. \TestCaseBatching performs poorly in extreme resource constraint conditions but effectively in high resource available environments, utilizing 80.65\% of baseline machines to maintain the actual average feedback time and 64.28\% to reach the plateaued feedback baseline. 
Both \DynamicBatching and \TestCaseBatching exhibit variable \ExecutionReduction. 
Notably, this study utilized simulated parallel test execution based on historical test results. 
Future research should confirm and extend these findings through experiments conducted in practical settings.


\section{Data availability} 
\label{sec_data}
The data for \Ericsson is not available as this is a proprietary project and cannot be released due to confidentiality. The data for Chrome is publicly available from Fallahzadeh~\etal~\cite{fallahzadeh_impact_2022}. 
We provide our scripts used to simulate the batching algorithms for Chrome at \url{https://github.com/emadfallahzadeh/BatchTesting}. 


\begin{acks}
   We acknowledge the support of the Natural Sciences and Engineering Research Council of Canada (NSERC) Discovery Grant and Concordia University FRS Funding. 
\end{acks}

\balance
\bibliographystyle{ACM-Reference-Format}
\bibliography{bibliography}
\end{document}